\documentclass[aps,prb,twocolumn,superscriptaddress,amsmath,amssymb]{revtex4-1}
\usepackage[utf8]{inputenc}
\usepackage{graphicx}
\usepackage{hyperref}
\hypersetup{colorlinks=true}
\usepackage{xcolor}
\normalsize
\usepackage{booktabs}
\usepackage{multirow}

\begin{document}

\title{LATTE: an atomic environment descriptor based on Cartesian tensor contractions}
\author{Franco Pellegrini}
\affiliation{Scuola Internazionale Superiore di Studi Avanzati, Via Bonomea 265, I-34136 Trieste, Italy}
\author{Stefano  de Gironcoli}
\affiliation{Scuola Internazionale Superiore di Studi Avanzati, Via Bonomea 265, I-34136 Trieste, Italy}
\affiliation{CNR-IOM DEMOCRITOS, Istituto Officina dei Materiali, Trieste, Italy.}
\author{Emine K\"{u}\c{c}\"{u}kbenli}
\affiliation{NVIDIA Corporation, Santa Clara, CA, USA}

\date{\today}
\begin{abstract}
We propose a new descriptor for local atomic environments, to be used in combination with machine learning models for the construction of interatomic potentials.
The Local Atomic Tensors Trainable Expansion (LATTE) allows for the efficient construction of a variable number of many-body terms with learnable parameters, resulting in a descriptor that is efficient, expressive, and can be scaled to suit different accuracy and computational cost requirements.
We compare this new descriptor to existing ones on several systems, showing it to be competitive with very fast potentials at one end of the spectrum, and extensible to an accuracy close to the state of the art.
\end{abstract}
\maketitle

\section{Introduction}\label{s:intro}
Machine learned interatomic potentials (MLIPs) have been proposed in recent years as an effective method to achieve \textit{ab initio} accuracy in the simulation of specific materials or molecular systems at a cost comparable with that of empirical potentials~\cite{Behler16,Schmidt19,Unke21,Kulik22}.
The common idea behind all MLIPs is the introduction of a flexible functional form with a large number of parameters, which can be optimized over several configurations of a given physical system to reproduce known observables---most commonly total energies---given the atomic configurations.
In the simplest setup, the input to the model is some form of \textit{descriptor} of the local environment (up to a certain cutoff) which is invariant under translations, rotations and permutations of equivalent atoms. Processing these atomic descriptors through different machine learning architectures (possibly having the atoms interact with each other) leads to some predicted local energies, which should sum to the total energy.
Within this scheme, and for a given number of training examples, the accuracy of the model depends on the quality of the descriptor and the architecture of the machine learning model.
Within a given architecture, and when comparing different ones, it is often the case that more accurate---or data efficient---models are also more computationally expensive at inference time, leading to a family of models with a trade-off between accuracy and efficiency.

Our new descriptor, called Local Atomic Tensors Trainable Expansion or LATTE, shares ideas with many existing descriptors, while proposing some new improvements to maximize its expressivity in a form which can be captured by a simple machine learning optimizer.
It is based on the construction of Cartesian tensors representing sum over spatially selected neighbors, and their contraction to obtain $n$-body scalar terms with favorable scaling. The use of learnable localized radial functions allows for the optimization of terms depending on a limited number of neighbors, which can be more easily fitted by the subsequent model.
In this work, we propose the use of LATTE in combination with atomic neural network, in the form of atomic multilayer perceptrons (MLPs), one of the simplest and most common machine learning models.
The LATTE descriptor is implemented in PANNA~\cite{PANNA1,PANNA2}, a software package for the creation of MLIPs. The new descriptor and neural network model are implemented in a new and efficient version of the code written in JAX~\cite{jax2018github}. The trained models can then be used to run simulations in ASE~\cite{ASE}, JAX-MD~\cite{jaxmd}, and LAMMPS~\cite{LAMMPS}, with good performance especially on GPUs.

In the next subsection, we will present in more detail the main descriptors that share similarities with LATTE. In the following section, we will describe the structure of LATTE in more details, before presenting its performance on several existing datasets.

\subsection*{Related work}\label{ss:related}
Many different functional forms have been proposed for local atomic environment descriptors.
While we refer the reader to one of the many reviews on the subject~\cite{Musil21} for a comprehensive overview, in this section we will mention the descriptors that are closest in functional form, accuracy or performance to LATTE.
Unless specified differently, we will consider functions centered around a given atom and describing its environment up to a given cutoff as a function of the position of its neighbors, i.e.\ all other atoms within the cutoff. The descriptor will in general be an array composed of a fixed number of components, or bins, to be used to fit a local atomic energy.

One of the first local atomic environment descriptors, typically referred to as Atom-Centered
Symmetry Functions (ACSF), was proposed by \citet{BP1}. This descriptor is composed of two sets of terms: 2-body terms based on a sum over neighbors and 3-body terms based on a sum of pairs of neighbors. In both cases the neighbor distance is sampled through Gaussians centered at fixed distances, and for the 3-body terms the angle between the neighbors is sampled through a cosine function of varying width, or center~\cite{ANI}. A further envelop function ensures continuity at the cutoff. Different species (or pairs of species for the 3-body terms) contribute to different elements of the descriptor.
This functional form is quite flexible and it has been successfully employed for the description of several systems~\cite{BP2, Behler21}, typically in combination with atomic neural networks consisting of species specific MLPs with a few hidden layers. However, this descriptor is not the most computationally efficient: it can lead to relatively large arrays, especially as the number of bins for the 3-body part scales quadratically with the number of atomic species.
Moreover, the computation of each 3-body term scales quadratically with the number of neighbors, which can be a rather large value especially for solids when considering a large cutoff. 
Finally, it has been pointed out~\cite{Pozdnyakov20} that limiting a descriptor to 3-body terms could lead to the same output for different environments, making the construction of an accurate MLIP impossible.

A more efficient functional form was proposed by Shapeev~\cite{MTP1, MTP2} in the Moment Tensor Potentials (MTP). In this descriptor, the positions of the neighbors are used to build tensor quantities, which are then contracted to obtain scalar elements. For each tensor, all neighbors are summed over, with a weight depending on the neighbor species and the neighbor distance, typically through a linear combination of Chebyshev polynomials.
In practice, this expansion is typically limited to a small number of tensor contractions and radial polynomials, and the potential is fitted through a single linear layer, for a very fast potential.
This functional form has the advantage to allow for an expansion in an increasing number of bodies, while retaining a linear scaling in the number of neighbors. Moreover, the descriptor size does not increase with the number of species considered.
However, the extremely compact form of the descriptor, and the use of radial polynomials potentially involving all atoms within the cutoff, can lead to a very complex dependence of the energy on the descriptor, which can be hard to fit with a single linear layer.
Moreover, the choice to limit the radial basis to very few elements means that the descriptor cannot be made more informative without resorting to interactions involving a very high number of bodies.
For this reason, this approach is typically used to obtain very fast potentials with moderate accuracy, rather than trading some of the computational cost for higher precision.

A similar structure, but on a different basis, can be found in the Atomic Cluster Expansion (ACE) descriptors~\cite{ACEDrautz, ACELinear}. In this case, a sum over neighbors projected on a spherical harmonics basis leads to multidimensional objects that are contracted with the appropriate Clebsch-Gordan coefficients to obtain scalar contributions. Like MTP, the contributions are weighted with some radial functions, which in general extend to all distances until the cutoff. However, species are handled separately in this case, with each species combination contributing to a different bin.
The choice of radial functions and spherical harmonics gives rise to contributions of increasing body order, with a size that increases in this case with a power of the number of species. However, the computational cost remains linear in the number of neighbors.
Like MTP, ACE descriptors are typically used in combination with a single linear layer. The larger basis, also considering the large number of different species combinations for high body order, has led to very accurate potentials~\cite{PACE}, with good computational efficiency.

Recently, a new efficient descriptor was proposed as the Ultra Fast (UF) potential~\cite{UF1}. The simplest form of this potential consists of 2-body contributions, writing the total energy as a linear combination of basis elements depending on pair distances. These basis elements are written as cubic splines, having a simple functional form and compact spatial support. 
Better accuracy can be reached by adding 3-body terms: all atoms triples within the cutoff contribute a linear combination of the product of splines of their 3 distances. While this is equivalent to a quadratic scaling with the number of neighbors, the small localized basis and linear fit keep the computational cost very low.

A different, very accurate approach is represented by equivariant graph neural networks~\cite{EGNN} (GNN). While many different architectures have been proposed, the common characteristic of these methods is the update of the local descriptor of one atom from those of its neighbors, through some form of message passing. 
In particular, equivariant GNNs based on spherical harmonics~\cite{NequIP,MACE} build local descriptors in a way similar to ACE, but keeping higher order irreducible representations and combining them to scalars only in the last layer. These models produce the most data efficient MLIPs, but they are often more expensive than the previous methods, due to the several operations required for each round of message passing, and its less parallelizable form.
A special case in this category is represented by Allegro~\cite{Allegro}, an architecture which shares many similarities with these models, but modifies the message passing to be completely local.

\section{Results}\label{s:results}
In this section, we will present the LATTE descriptor in more detail, and we will show its performance in different settings.
In particular, we will show the efficiency for very small descriptors as compared to some of the fastest models on a Tungsten crystal dataset (\ref{ss:W}), we will compare the data efficiency for larger models on a standard benchmark of small molecules (\ref{ss:rMD17}), and finally we will show the accuracy and model capacity in an application to a very large organic dataset with multiple atomic species (\ref{ss:SPICE}).

\subsection{The model}\label{ss:model}
The LATTE descriptor is based on the construction of contributions representing an atomic environment as a sum over neighbors of radial functions and tensors of the relative neighbor positions. The tensors are then contracted to obtain scalar quantities, similarly to the MTP and ACE approaches. Unlike those approaches, however, we employ a trainable radial basis with limited radial support, leading to a larger variety of terms for the same tensorial contraction, each involving a smaller number of neighbors. This larger basis, coupled with the larger number of parameters of atomic neural networks, leads to a greater model expressiveness.

More in detail, we start by defining the local tensors around atom $i$ as:
\begin{equation}
    \displaystyle A_{i,u}^{\alpha_1\ldots\alpha_p}=\sum_{j\in\mathcal{N}(i)}\sigma_u(s_i,s_j)\,f_{u,s_i}(\left|\vec{r}_{ij}\right|)\,\,
    \hat{r}_{ij}^{\alpha_1} \otimes\ldots\otimes \hat{r}_{ij}^{\alpha_p},
\end{equation}
where $j$ runs over all the neighbors $\mathcal{N}(i)$ of atom $i$, $s_j$ represents the species of atom $j$, $\vec{r}_{ij}$ is the vector from $i$ to $j$, and $\hat{r}_{ij}=\vec{r}_{ij}/\left|\vec{r}_{ij}\right|$ the corresponding unit versor.
The superscripts $\alpha_1\ldots\alpha_p$, for a tensor of order $p$, correspond to the tensor components, coming from the outer product of $p$ copies of the versors $\hat{r}_{ij}$. The index $u$ identifies one specific instance of this tensor product, corresponding to a specific set of parameters in the functions $f_u$ and $\sigma_u$.

In the following, the species function $\sigma_u$ will simply be a learnable scalar for each possible value of species $s_i$ and $s_j$, i.e.\ for a system with $S$ possible species, we define a $S\times S$ matrix $\Sigma^u$ for each term and $\sigma_u(s_i,s_j)=\Sigma^u_{s_i,s_j}$.
For the radial function $f_{u,s_i}$, we employ a localized function with a computationally efficient functional form:
\begin{equation}
    \displaystyle f_{u,s_i}(r)=\frac{1}{r_{u,s_i}^2}\left[\mathrm{ReLU}\left(1-\left(\frac{r-r_{u,s_i}}{w_{u,s_i}}\right)^2\right)\right]^3,
\end{equation}
where $r_{u,s_i}$ and $w_{u,s_i}$ are the learnable parameters (one for each bin $u$ and species $s_i$) and $\mathrm{ReLU}$ represents a Rectified Linear Unit, i.e.\ $\mathrm{ReLU}(x)=x$ for positive $x$, and $0$ otherwise. This function is centered around $r_{u,s_i}$ and different from zero only in $[r_{u,s_i}-w_{u,s_i},r_{u,s_i}+w_{u,s_i}]$, so that only some of the neighbors can contribute to this term. Since the function and its first and second derivative are zero at the boundary, the descriptor is smooth at the cutoff $R_c$ as long and we ensure that $r_{u,s_i}+w_{u,s_i}<R_c$ during training. Moreover, the prefactor $1/r_{u,s_i}^2$ counterbalances the tendency for neighbors to grow quadratically for a given interval at increasing distance from the center.

From $N$ tensor terms, we can build a $N+1$-body scalar term by contracting the indices:
\begin{equation}
    \displaystyle B_{i,u} = \sum_{\alpha_1,\ldots,\alpha_m} 
    \prod_{k=1}^N A_{i,u_k}^{\alpha^k_1\ldots\alpha^k_{p_k}},
\end{equation}
where it is implied that each of the indices $\alpha_1,\ldots,\alpha_m$ is found twice among the $\alpha^k_j$, i.e.\ each component of a tensor is contracted with a single other component.
In this notation, the indices $u_k$ indicate that each term $A_{i,u_k}$ contains its own learnable parameters, while $u$ is the overall index for the final scalar at bin $u$.

It should be noted that the 2-body term can be treated as a slightly special case, where $B_{i,u}$ corresponds to a single $A_{i,u}$ with no indices, which is a simple scalar function of the neighbor species and distances.

To obtain an interatomic potential, the descriptor is fed to a MLP of a few layers, with a nonlinearity in all but the last layer (a Gaussian in these exmaples).
In the general case the network weights are shared between atoms of the same species, as commonly done for ACSF descriptors, as are the descriptor parameters. For an even more computationally efficient model, the descriptor, and some of the MLP layers, can also have the same weights shared for all atoms, irrespective of the species.
The descriptor and neural network optimization is implemented in a new extension of the PANNA code~\cite{PANNA1, PANNA2}.

While this descriptor shares its underlying structure with ACE and MTP---and the cartesian tensor expansion with the latter---some of the new design choices increase its expressive power. 
The choice of a localized radial function, borrowed from the ACSF-style descriptors, decreases the number of neighbors that contribute to the value of a single bin. In our experience, this allows the subsequent model to more accurately fit single features, leading to better generalization with the same body order.
As a natural consequence, the possibility to vastly increase the number of elements considered for a given body order (each element having learnable parameters) further increases the expressive power with a modest increase in computational cost.
It has to be noted that, as a design choice, both ACE and MTP propose heuristics to solve the problem of trade-off between increasing the radial basis elements and increasing the angular basis or the body order.
In the case of LATTE, we leave this choice to the users, allowing them to select custom tensor contractions and an arbitrary number of elements for each of them.
In Sec.~\ref{ss:rMD17}, we will show how the accuracy of the model can often be increased by adding tens of elements for the same contraction, before they become superfluous and it becomes more beneficial to move to a new contraction. Since the cost of each element scales linearly in the body order, and as a power of the tensor indices, using the largest number of elements for each contraction is often beneficial.
Mixing all atomic species in each term of the descriptor (as in the MTP case) means that a small descriptor can be used even for a large number of possible species. Conversely, the descriptor size can be manually increased for more varied datasets. 
Lastly, the use of a neural network model rather than a single linear layer further increases the number of parameters and the expressive power of the model, while only increasing computational cost moderately as compared to higher order descriptor elements.

As a notation remark, in the following we denote terms by a string composed of the number of elements, followed by a comma separated list of indices: each set of indices refers to a single tensor, and same indices are summed over in the final descriptor.
As a practical example, the string 100($\alpha$,$\beta$,$\alpha\beta$) refers to 100 elements of the 4-body terms defined as:
\begin{equation}
    \displaystyle B_{i,u} = \sum_{\alpha,\beta} 
    A_{i,u_1}^{\alpha}A_{i,u_2}^{\beta}A_{i,u_3}^{\alpha\beta}.
\end{equation}
In the special case of 2-body terms, where no index is contracted, an empty parentheses notation is used, e.g.\ 20() for 20 2-body bins.

\subsection{Comparison with fast models: W crystals}\label{ss:W}

In this section, we show the computational efficiency of LATTE for small descriptor sizes by training potentials on a small dataset of Tungsten crystals presented in \citet{Szlachta14}, and comparing them to the performance of UF potentials~\cite{UF1}.

\begin{table}
    \centering
    \caption{Validation RMSE in energy and forces for different potentials trained on a Tungsten dataset~\cite{Szlachta14}. The last column reports the time in ms for one MD step of a 128 atoms cell on a single CPU core.}
    \label{tab:W}
    \begin{tabular}{cccc}
    \toprule
        Model & E RMSE & F RMSE & Time/step\\
         & [meV/atom] & [meV/\AA] & [ms]\\
    \midrule
        UF$_2$ & 26.6~\cite{UF1} & 387~\cite{UF1} & 0.3\\
        UF$_{2,3}$ & 5.1~\cite{UF1} & 152~\cite{UF1} & 4.0\\
        LATTE & 5.6~~ & 149~~ & 2.4\\
    \bottomrule
    \end{tabular}
\end{table}

To reproduce the settings of \citet{UF1}, we consider a small training set of 1939 configurations, and keep the rest for validation.
We consider a small descriptor up to 3 bodies: 18 2-body elements and 14 3-body contractions of type ($\alpha,\alpha$).
We keep atoms up to a 5~\AA\ cutoff and we consider a very small network with a single hidden layer of size 64.
We optimize the model over batches of 50 examples through the Adam~\cite{Adam} optimizer with learning rate $10^{-4}$ for $2\cdot10^6$ steps, and then reduce it exponentially to $10^{-6}$ over the next $2\cdot10^6$ steps. We use a small L1 regularization $10^{-4}$.
The cost function is the sum of the quadratic error in energy per atom, with coefficient 200, and the quadratic error in force per component.

Table~\ref{tab:W} shows the root mean square error (RMSE) in energy per atom and force components for the LATTE model as well as the 2- and 3-body UF models. We also report the single core timings per step to perform NVE molecular dynamics (MD) for a cell of 128 atoms on a Intel Core i7-9850H CPU. While this is not the most general benchmark, it is in line with the original publication, and is aimed at showcasing the performance for very limited computational resources.
The UF simulations are performed through the LAMMPS~\cite{LAMMPS} plugin, with a pretrained potential on our hardware. It has to be noted that the original publication was reporting even faster performance with a tabulated version of the potential, however for this comparison the commonly available non-tabulated implementation has been used.
The LATTE simulations are performed through JAX-MD~\cite{jaxmd}.
We can see that the accuracy and performance of the LATTE model are comparable to those of the 3-body UF potential.

\subsection{Small molecules: rMD17}\label{ss:rMD17}
In recent years the dataset known as rMD17~\cite{rMD17} has been used in several publications as a benchmark for the data efficiency of interatomic potentials.
It consists of configurations of 10 small organic molecules, with energies and forces computed in DFT with a tight convergence threshold.
To test data efficiency, we follow the commonly used fitting setup consisting in training a model on just 1000 randomly selected configurations per molecule.

\begin{figure}[!h]
    \centering
    \includegraphics[width=\linewidth]{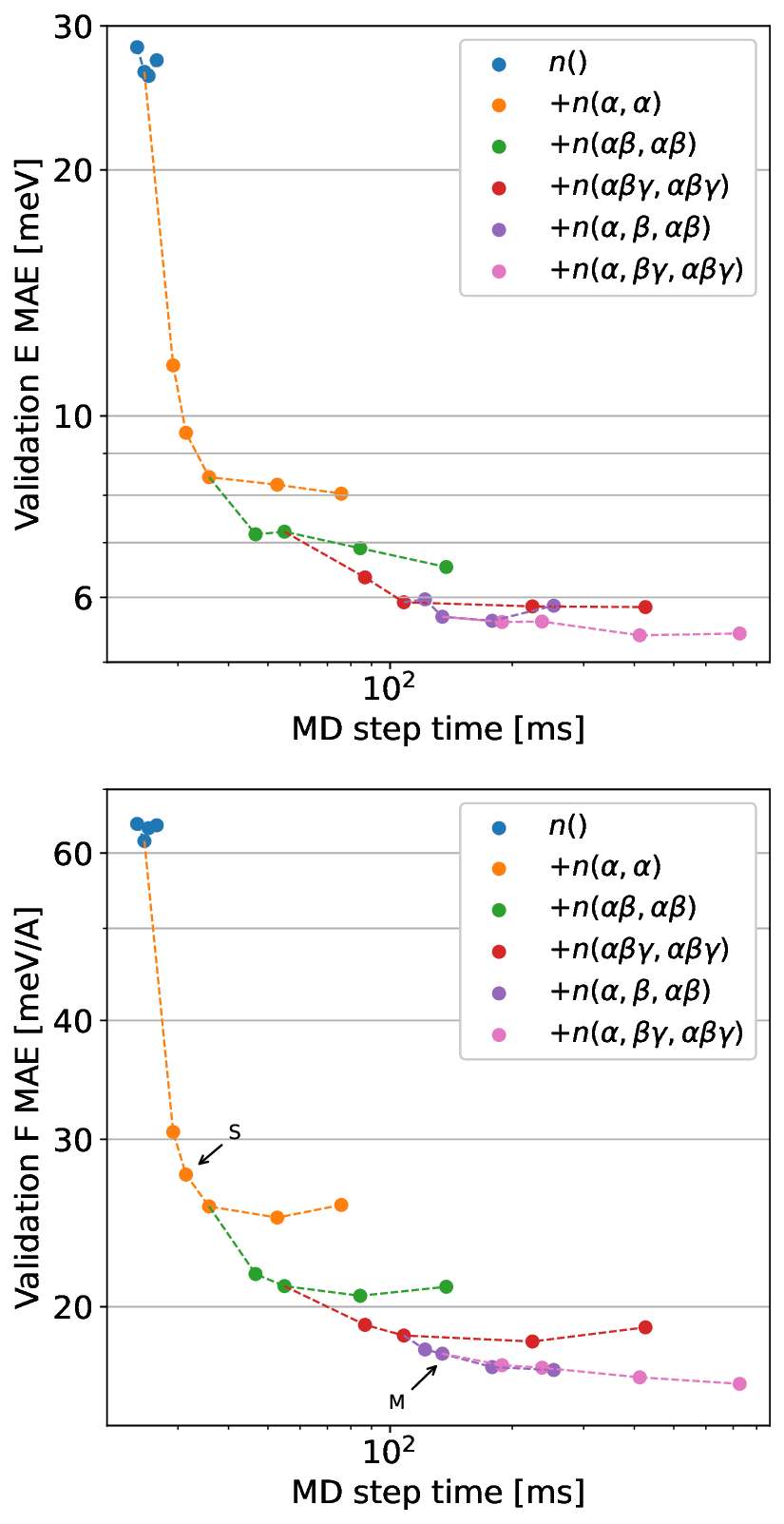}
    \caption{Validation mean absolute error in energy per molecule (top) and forces per component (bottom) for different descriptors trained on aspirin configurations, as a function of the time needed to run a MD step of 1000 molecules.
    Each color represents the addition of a new contraction, with increasing number of terms. The starting descriptors only have 2-body terms, indicated as $n$() in the legend, and successive sets start from the previous descriptor (connected with a dashed line) and add an increasing number of terms of the contraction indicated in the legend.
    The points indicated by arrows and labeled S and M are those used in Table~\ref{tab:rmd17}.}
    \label{fig:rMD17Scal}
\end{figure}

For ease of comparison, throughout this section we use the same network and training procedure. Specifically, we employ an atomic network for each species with 2 hidden layers of sizes 256 and 128, with Gaussian nonlinearities.
We train all models for $3\cdot10^6$ minibatches of 10 configurations, through the Adam~\cite{Adam} optimizer with a constant learning rate $5\cdot10^{-5}$ for the first half of the training, followed by an exponential quench to $5\cdot10^{-7}$. We employ a small L1 normalization $10^{-3}$. The loss function is the sum of the quadratic energy loss on the whole molecule, and the mean quadratic force loss per component, with a prefactor 500.
As commonly found when training MLIPs, we find a trade-off between energy and forces accuracy. This prefactor and training parameters slightly favor the force components of the potential.
While varying the terms in the descriptor, we keep a cutoff of 5~\AA\ for all of them, and we do not optimize the $w_u$ values, but keep them fixed at 1~\AA.

We first show how the model accuracy varies by increasing the number and variety of elements in the atomic descriptor.
In Fig.~\ref{fig:rMD17Scal} we show the validation mean absolute error (MAE) in energy per molecule and forces per component for descriptors with an increasing number of terms trained on 1000 aspirin configurations, as a function of the inference time. Inference was run on a single  NVIDIA A100 GPU and in order to show a realistic scaling (that can use the available resources) we report the time per step to simulate 1000 molecule, for a total of 21000 atoms.
Each color represents the addition of a new tensor contraction (see legend) with an increasing number of terms ranging from 30 to 1000. The dashed lines connect each new contraction to the number of elements used from the previous contraction. See the SM for an explicit list of all contractions used.

We can see how increasing the number of terms for each contraction initially leads to a consistent improvement in accuracy, before resulting in smaller improvement or an overfit, especially for the force components.
Conversely, adding new contractions is usually beneficial, although the improvement becomes progressively more modest as we go to more complex contractions, with an increasing computational cost.

\begin{table}[!t]
\caption{
Mean absolute error in energy (meV on the whole molecule) and forces (meV/\AA\ per component) of different models trained on 1000 configurations from each molecule in the rMD17 dataset~\cite{rMD17}.
The first two columns are taken from the literature, and the following two are LATTE models with differently sized descriptors (see text).
In the last column we report the state of the art (SOTA), i.e.\ the best result found for any model, giving priority to the force error, and the respective reference.\\
}
\centering
\label{tab:rmd17}
\resizebox{0.48\textwidth}{!}{
\begin{tabular}{@{}llccccc@{}}
\toprule
&& ACSF~\cite{PANNA2}& ACE~\cite{ACELinear} & LATTE & LATTE &{SOTA}\\
&& && small & medium &\\
\midrule
\multirow{2}{*}{\textbf{Aspirin}}
& E & 10.6& 6.1 & 9.5 & 5.7 & 2.2~\cite{MACE}\\
& F & 32.9& 17.9 & 27.5 & 17.8 &6.6\\
\midrule
\multirow{2}{*}{\textbf{Azobenzene}}
& E & 5.8& 3.6 & 7.0 & 3.3 & 1.2~\cite{Allegro}\\
& F & 18.4& 10.9 & 19.5 & 10.4 & 2.6\\
\midrule
\multirow{2}{*}{\textbf{Benzene}}
& E & 1.0& 0.04 & 0.73 & 0.42 &0.3~\cite{Allegro}\\
& F & 5.4& 0.5 & 4.6 & 1.1 & 0.2\\
\midrule
\multirow{2}{*}{\textbf{Ethanol}}
& E &2.9 & 1.2 & 2.5 & 1.2 & 0.4~\cite{MACE,Allegro}\\
& F & 16.5& 7.3 & 14.2 & 6.3 & 2.1\\
\midrule
\multirow{2}{*}{\textbf{Malonaldehyde}}
& E & 4.0& 1.7 & 3.9 & 1.7 & 0.6~\cite{Allegro}\\
& F & 24.3& 11.1 & 20.4 & 10.6 &3.6 \\
\midrule
\multirow{2}{*}{\textbf{Naphthalene}}
& E & 3.0& 0.9 & 3.6 & 1.6 & 0.2~\cite{Allegro}\\
& F & 13.2& 5.1 & 15.1 & 6.3 &0.9\\
\midrule
\multirow{2}{*}{\textbf{Paracetamol}}
& E & 6.3& 4.0 & 5.8 & 3.5 & 1.3~\cite{MACE} \\
& F & 22.0& 12.7 & 20.9 & 12.6 &4.8\\
\midrule
\multirow{2}{*}{\textbf{Salicylic acid}}
& E & 4.1& 1.8 & 4.1 & 2.4 & 0.9~\cite{Allegro} \\
& F & 19.4& 9.3 & 18.3 & 9.6 &2.9\\
\midrule
\multirow{2}{*}{\textbf{Toluene}}
& E &3.9& 1.1 & 3.6 & 1.5 & 0.5~\cite{MACE}\\
& F & 15.9& 6.5 & 15.3 & 6.6 & 1.5\\
\midrule
\multirow{2}{*}{\textbf{Uracil}}
& E & 2.4& 1.1 & 3.2 & 1.1 & 0.6~\cite{Allegro}\\
& F & 13.7& 6.6 & 12.2 & 5.7 & 1.8 \\
\bottomrule
\end{tabular}
}
\end{table}

Having seen the effects on accuracy and computational cost of increasing the descriptor size, we now compare two different models to some of those reported in the literature trained on 1000 configurations for each of the 10 molecules in the dataset.
For each molecule, we use exactly the same parameters as the aspirin experiments, without further optimizing them.
We consider 2 descriptors: a ``small'' one just 150 elements up to 3-body with a single index, and a ``medium'' one with 850 elements up to 3-body with 3 indices and 4-body with 2 indices
(the exact composition is specified in the SM, the arrows in Fig.~\ref{fig:rMD17Scal} indicate these models as S and M).

We can see how the small descriptor is generally in line with, or better than, the ACSF results. This is interesting as the fitting model is exactly the same, and the LATTE descriptor is considerably smaller, and shows better scaling with the nearest neighbors.
The middle descriptor is mostly in line with the ACE results, with the exception of naphthalene, where the ACE potential used a larger cutoff, and benzene.
It is however still about a factor 3 from the ``state of the art'' (SOTA) results (with the exception of naphthalene, where a larger cutoff was also used, and benzene), but the equivariant graph convolutional networks employed to reach this accuracy have a higher computational cost.

\subsection{A large dataset: SPICE}\label{ss:SPICE}
To showcase the capacity of the LATTE descriptor, we train model on a very large and challenging dataset.
The Small-molecule/Protein Interaction Chemical Energies (SPICE) dataset~\cite{SPICE} was recently proposed as a dataset relevant to simulating drug-like small molecules interacting with proteins. It consists of 1.1 million conformations, including small molecules, dimers, dipeptides, and solvated amino acids, for a total of 15 atomic species. 
The presence of so many species, in particular, limits the applicability of any approach with an unfavorable scaling with respect to the number of elements, while the variety of chemical environments requires a flexible fitting model.
On the other hand, a fast model trained on this dataset is relevant to the simulation of many system of biological interest.

\begin{table}[!t]
    \caption{Comparison of different models trained on the SPICE dataset.
    The first 2 column indicate: whether the model used is small (S) or large (L) and whether the descriptor and early weights were species-specific (S) or common (C). 
    For each models we then report the MAE on energies per atom and forces per component, and the time required for an MD simulation per step per atom on a A100 GPU.}
    \centering
    \begin{tabular}{ccccc}
    \toprule
      \multirow{2}{*}{Model} & \multirow{2}{*}{Weights} & 
      E MAE & F MAE & Time/step/atom\\
       & & [meV/atom] & [meV/\AA] & [$\mu$s] \\
    \midrule
       S & C & 22.4 & 66.9 & 1.7 \\
       S & S & 16.4 & 53.9 & 4.1 \\
       L & S & 13.7 & 45.8 & 8.5 \\
    \bottomrule
    \end{tabular}
    \label{tab:SPICE}
\end{table}

For our fits, we use version 1.1.3 of the dataset.
We follow the original authors and remove all configurations with a large force component: in line with \citet{AllegroSPICE} we set the cutoff at 0.25~Ha/bohr. We select 90\% of the configurations for training, 5\% for validation and 5\% for test, trying to match this ratio as closely as possible for each set of configurations in the dataset.
We consider two different models: a smaller one with a descriptor of size 256 and a larger model with a descriptor of 512 terms, both with hidden layers of sizes 256, 256 and 128
(see SM for a complete description). 
For the small model, we also consider the possibility of using the same descriptor, and first 3 weights layers as common for all atom, regardless of atomic species.
In all cases the cutoff is set at 4~\AA.
We train on batches of size 100 with Adam~\cite{Adam} optimizer with a learning rate $10^{-4}$ for $6\cdot10^7$ steps then exponentially quenched to $10^{-6}$ over $2\cdot10^7$ steps. We apply a small L1 regularization coefficient $10^{-4}$. The loss function is the sum of the quadratic error per atom, with a coefficient $10^2$, and the quadratic error per force component, with a coefficient $10^3$.

In Table~\ref{tab:SPICE} we report the validation MAE in energy and forces for various descriptors and models.
We use each potential to perform MD simulations through our JAX-MD plugin of the Dihydrofolate Reductase (DHFR) system from the AMBER20 benchmark~\cite{Amber}, a large molecule in a water box for a total of 23k atoms. We report the time per step per atom on a single A100 GPU in the same table.

To put our results in context, the equivariant model of \citet{AllegroSPICE} is trained on the same exact data and reports a force MAE of 25.7 meV/\AA. They perform MD on the DHFR system and report performance for various number of GPUs (a minimum of 4), corresponding to a minimum time per step per atom per GPU of around 10~$\mu$s.
Another model trained on similar data is presented in \citet{MACESPICE}: the starting dataset is the same, but it is refined and augmented with some extra data. The validation error is thus not directly comparable, and it is reported as RMSE on different data splits, but just as a baseline it varies between 10 and 30~meV/\AA\ for the largest model and between 25 and 60~meV/\AA\ for the smaller one.
MD performance is reported for the smallest model simulating water with varying number of atoms and corresponds to around 35~$\mu$s per step per atom on a single A100 GPU, while multiple GPUs performance is much lower due to the burden of message passing.
A model with a more comparable architecture, presented in \citet{DeepMD2}, has an attention-based version modified to handle a large number of atomic species which was also trained on the SPICE dataset. The training was performed on 95\% of the whole dataset, to a validation RMSE of 78~meV/atom in energy and 233~meV/\AA\ in forces. MD performance are reported for equivalent models on a A100 GPU around 20~$\mu$s per step per atom.

Our simple model can thus be trained on this complex dataset to an accuracy within a small factor of the best models, while being faster at inference time. As is usual for these models, the speed can be further increased at a cost for the accuracy.

\section{Conclusion}\label{s:conclusion}
We have presented LATTE, a new descriptor for the construction of MLIPs.
We have shown that models based on LATTE and MLPs of different sizes perform well for different datasets and accuracy targets: small descriptors and shallow models can compete with the fastest MLIPs, while larger models can reach an accuracy surpassed only by larger GNNs. Moreover, we have shown that LATTE models can be trained on very large datasets with several atomic species and achieve a good accuracy while being computationally competitive even for the simulation of very large systems.

It has to be noted that while in the very small dataset regime (see e.g.\ Sec.~\ref{ss:rMD17}) even simple models can overfit to the training set and validation accuracy is limited by the inherent architectural bias, for very large datasets the capacity of the model becomes more relevant and as we show in Sec.~\ref{ss:SPICE} even simple but large architectures can be trained to a high accuracy.
This makes LATTE a very valuable approach, given the recent focus on very large dataset and general models, and its proven capacity to handle complex datasets and multiple species while retaining high efficiency.

In particular, even the simplest form of weights sharing between different species seems to further increase the performance of our model.
An extension of this, where atomic species are embedded based on their physical properties, is a natural evolution of this approach that will be the subject of future work.

Given the vast landscape of existing models, we believe that LATTE constitutes a valid alternative to be considered when building MLIPs for different systems.
Its capacity to handle multiple species and the possibility to gradually improve its accuracy and handle even very varied datasets, while not unique, make it a strong contender in many real world applications.

\section{Data and code availability}
All the datasets used for training are publicly available.
We release the code for the descriptor, new training in JAX and the MD plugins as part of the PANNA package~\cite{PANNAurl}.

\bibliography{LATTE}

\clearpage
\section*{Supplementary Material -- Models details}\label{sm:details}
\subsection{Models used in Sec.~IIC}
For each of the ``series'' shown in Fig.~1 (points of the same color), we report in Table IV below the type of terms added and the multiplicities explored.
The bold number is the one kept as a starting point for the next series.

\begin{table}[!b]
\caption{}
\centering
\label{tab:rmd17Curves}
\resizebox{0.4\textwidth}{!}{
\begin{tabular}{lcl}
\toprule
Term&&Sizes\\
\midrule
()&&30, \textbf{50}, 100, 200\\
($\alpha,\alpha$)&&50, 100, \textbf{200}, 500, 1000\\
($\alpha\beta,\alpha\beta$)&&100, \textbf{200}, 500, 1000\\
($\alpha\beta\gamma,\alpha\beta\gamma$)&&
100, \textbf{200}, 500, 1000\\
($\alpha,\beta,\alpha\beta$)&&100, \textbf{200}, 500, 1000\\
($\alpha,\beta\gamma,\alpha\beta\gamma$)&&
100, 200, 500, 1000\\
\bottomrule
\end{tabular}
}
\end{table}

For the results reported in Table II of the main text, the ``small'' descriptor  has a structure 50(), 100($\alpha,\alpha$); while the ``medium" model has structure 50(), 200($\alpha,\alpha$), 200($\alpha\beta,\alpha\beta$), 200($\alpha\beta\gamma,\alpha\beta\gamma$), 200($\alpha,\beta,\alpha\beta$).

\subsection{Models used in Sec.~IID}
For the results reported in Table III of the main text, the ``small'' model uses a descriptor 56(), 100($\alpha,\alpha$), 100($\alpha\beta,\alpha\beta$); the ``large'' model 56(), 128($\alpha,\alpha$), 128($\alpha\beta,\alpha\beta$), 128($\alpha\beta\gamma,\alpha\beta\gamma$), 128($\alpha,\beta,\alpha\beta$).
\end{document}